\PassOptionsToPackage{switch}{lineno} 
\documentclass[conference]{IEEEtran}
\usepackage{fancyhdr}
\IEEEoverridecommandlockouts
\usepackage{cite}
\usepackage{xspace}
\usepackage{amsmath,amssymb,amsfonts}
\usepackage{algorithmic}
\usepackage{graphicx}
\usepackage{textcomp}
\usepackage{subfig}
\usepackage{makecell}
\usepackage{array}
\newcolumntype{P}[1]{>{\centering\arraybackslash}p{#1}}
\newcolumntype{M}[1]{>{\centering\arraybackslash}m{#1}}
\usepackage{booktabs}
\usepackage[most]{tcolorbox}
\usepackage{ragged2e}
\usepackage{enumitem}
\usepackage{comment}
\usepackage{xurl}

\usepackage{pifont}
\newcommand{\cmark}{\ding{51}}
\newcommand{\xmark}{\ding{55}}

\usepackage{xcolor}
\def\BibTeX{{\rm B\kern-.05em{\sc i\kern-.025em b}\kern-.08em
    T\kern-.1667em\lower.7ex\hbox{E}\kern-.125emX}}
    
\newcommand{\todo}{{\color{red}\bfseries TODO}}

\begin{document}

\title{\approach: Leveraging LLM Agents to Compose, Catalog, and Deploy Reproducible Workflows\\
}

\DeclareRobustCommand{\approach}{CURATE\xspace}
\newcommand{\Td}{$T_d$\xspace}
\newcommand{\Tr}{$T_r$\xspace}
\newcommand{\Tu}{$T_u$\xspace}
\newcommand{\Tc}{$T_c$\xspace}
\newcommand{\Ta}{$T_a$\xspace}

\newtcolorbox{promptbox}[1][Prompt]{
  colback=black!4, colframe=black!30, boxrule=0.4pt, arc=1pt,
  left=5pt, right=5pt, top=2pt, bottom=2pt, breakable,
  before skip=2pt, after skip=2pt,
  fontupper=\ttfamily\scriptsize\selectfont,
  title={#1},
  halign title=center,
  fonttitle=\ttfamily\small\bfseries,
  coltitle=black,
  colbacktitle=black!4,          
  titlerule=0.4pt,               
  titlerule style=black!30,
}

\newlist{promptitems}{itemize}{1}
\setlist[promptitems]{leftmargin=1.2em,topsep=1pt,itemsep=0.5pt,parsep=0pt,label=\textbullet}

\newtcolorbox{revisionbox}{
  colback=violet!5, colframe=violet!55!black, boxrule=0.4pt, arc=1pt,
  left=5pt, right=5pt, top=2pt, bottom=2pt, breakable,
  before skip=2pt, after skip=2pt,
  fontupper=\ttfamily\scriptsize\selectfont,
  title={Revisions},
  halign title=center,
  fonttitle=\ttfamily\small\bfseries,
  coltitle=violet!55!black,
  colbacktitle=violet!5,
  titlerule=0pt,
}

\makeatletter
\newcommand{\newlineauthors}{%
  \end{@IEEEauthorhalign}\hfill\mbox{}\par
  \mbox{}\hfill\begin{@IEEEauthorhalign}
}
\makeatother

\author{
\IEEEauthorblockN{Nolan Cutler}
\IEEEauthorblockA{\textit{School of EECS} \\
\textit{Oregon State University}\\
Corvallis, Oregon, USA\\
cutlern@oregonstate.edu}
\and
\IEEEauthorblockN{Chia-Chen Kuo}
\IEEEauthorblockA{\textit{School of EECS} \\
\textit{Oregon State University}\\
Corvallis, Oregon, USA\\
kuochi@oregonstate.edu}
\and
\IEEEauthorblockN{Nanda Velugoti}
\IEEEauthorblockA{\textit{School of EECS} \\
\textit{Oregon State University}\\
Corvallis, Oregon, USA\\
velugotn@oregonstate.edu}
\newlineauthors
\IEEEauthorblockN{Kathryn Newhart}
\IEEEauthorblockA{\textit{School of CBEE} \\
\textit{Oregon State University}\\
Corvallis, Oregon, USA\\
newhartk@oregonstate.edu}
\and
\IEEEauthorblockN{Renato Figueiredo}
\IEEEauthorblockA{\textit{School of EECS} \\
\textit{Oregon State University}\\
Corvallis, Oregon, USA\\
figueren@oregonstate.edu}
}


\maketitle
\thispagestyle{fancy}
\lhead{}
\rhead{}
\chead{}

\lfoot{\footnotesize{%
SC26 Workshops, November 15-20, 2026, Chicago, Illinois, USA
\newline
979-8-3195-1221-5/26/\$31.00 \copyright 2026 IEEE%
}}
\rfoot{}
\cfoot{}

\renewcommand{\headrulewidth}{0pt}
\renewcommand{\footrulewidth}{0pt}

\begin{abstract}
Agentic code generation has the potential to accelerate the development of computational workflows while also reducing barriers to entry. However, a key gap remains: existing coding agents focus on code generation and do not address the entire workflow lifecycle, including deployment and sharing. As a result, users develop and stitch modules independently while managing deployment on their own. To address this gap, we propose CURATE (Composition, User-in-the-loop, Reuse, and Automated Task Execution), a novel human-in-the-loop multi-agent system that uses LLM agents to manage and develop composable workflows across their entire lifecycle. A key feature of the system is a catalog that allows for the storage and reuse of modules across workflows. Module catalogs provide a foundation that can be expanded to support FAIR principles by facilitating the sharing and reuse of curated modules and subgraphs. We demonstrate the feasibility of our system with an initial prototype and 6 experiments. 
\end{abstract}


\section{Introduction}
\label{sec:intro}

Scientific workflows are a vital part of the scientific discovery process. They allow researchers to automate data ingestion and computational experiments, reproduce their results, and share their research artifacts with the broader scientific community~\cite{kepler, deelman-taxonomy}. However, workflows require significant effort and time to build. Users must develop familiarity with workflow management tools, source or write workflow steps by hand, and manage deployment infrastructure. These systems have a steep learning curve that hinders adoption and takes time away from other research activities~\cite{community-roadmap}. Opportunely, generative AI has shown promise in accelerating parts of the scientific software development process~\cite{llm-for-science, code-translation-science}. Large language models (LLMs) can generate code from natural language, and systems such as LLM4Workflow~\cite{LLM4workflow} demonstrate their ability to compose multi-step workflows. Recently, LLM agents have been utilized for their ability to autonomously plan tasks and invoke tools~\cite{toolformer, swe-agent, ReAct}, self-correct~\cite{reflexion}, and incorporate human expertise into their decision-making~\cite{hitl-agents}. Moreover, coding agents, such as Claude Code and Codex, have demonstrated a remarkable ability to generate, test, and execute code locally on a user's device, with human-in-the-loop (HITL) verification~\cite{claudecode, codex}. At the same time, initiatives such as the Workflow Community Initiative (WCI), WorkflowHub~\cite{workflowhub}, and CWL~\cite{cwl} have pushed to imbue workflows with FAIR (Findable, Accessible, Interoperable, Reusable) principles~\cite{FAIR}. Still, existing agent-driven workflow generation systems lack an end-to-end view of the scientific workflow lifecycle, from generating and composing workflow steps to deployment and monitoring to broader FAIR reuse and sharing of validated components. The contributions of our work are the design and proof-of-concept implementation of the core functionality of a novel multi-agent system for scientific workflow generation, incorporating existing components, as well as early experiments demonstrating its feasibility.

\section{Related Work}
\label{sec:related}

\begin{figure*}[t]
\centering
\includegraphics[width=\textwidth]{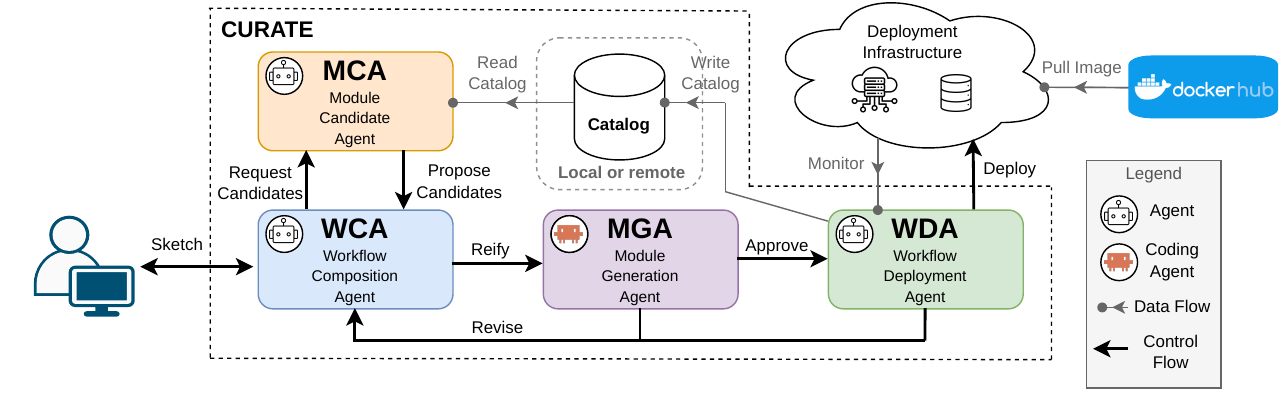}
\caption{Overview of the CURATE architecture. The user iterates with the multi-agent system through sketch $\rightarrow$ reify $\rightarrow$ test $\rightarrow$ deploy phases, with checkpoint gates allowing user verification and revision, and a catalog for code sharing and reuse.}
\label{fig:curate-architecture}
\end{figure*}

\textbf{Scientific Workflows}:
Scientific workflows describe processes orchestrated to achieve a scientific goal. They are typically expressed as directed graphs of computational tasks with data dependencies and executed by a workflow management system (WMS) that handles scheduling, reproducibility, data movement, and fault tolerance~\cite{ornl-terminology}. A wide range of WMSs exists, from high-performance computing (HPC) targeted systems to cloud approaches~\cite{ornl-terminology}. Our proposed architecture uses an intermediate workflow representation -- a skeleton -- that is not inherently tied to any particular WMS.

\textbf{Agents for Workflow Generation}: Coding agents, such as Claude Code and Codex, are LLM-enabled agents for code generation~\cite{claudecode, codex}. These systems integrate LLM inference, reasoning, and tool calling to autonomously iterate through code generation and execution, providing initial checks before human verification. Agents can use external tools and skills that can provide reusable instructions, scripts, and resources~\cite{toolformer, skills}. Thareja et al.~\cite{deelman} apply skills to Claude Code, Codex, and Kimi for workflow generation targeting Pegasus~\cite{pegasus}, demonstrating the potential for agentic workflow development. While sharing similarities, CURATE introduces a key novel capability: a catalog that supports storage, retrieval, and reuse of user-verified modules by the agentic system. Lee et al.~\cite{spreadsheet-llm} apply HITL and module reuse through Retrieval Augmented Generation; however, their system targets Excel and CSV spreadsheet data-cleaning tasks.

\textbf{FAIR and Workflow Cataloging}: Prior work has integrated FAIR principles into workflow development through workflow registries, platform-agnostic workflow languages, and workflow component search~\cite{workflowhub, cwl, workflow-search}. WorkflowHub~\cite{workflowhub} provides a registry of computational workflows for sharing and reuse. CWL~\cite{cwl} provides a partially platform-agnostic solution to workflow definition. Starlinger et al.~\cite{workflow-search} demonstrate the feasibility of similarity search for workflows, but do not provide a way to reuse components during workflow composition automatically. These projects are complementary to CURATE. WorkflowHub, for example, could serve as a foundation for creating pre-seeded catalogs of expert-vetted, auditable workflow components; CWL could be used as a target workflow representation to enable a wider range of WMS deployment options; and advanced catalog searching techniques could improve the accuracy of module sourcing.

\textbf{Commercial Science Assistants}: Google AI Co-Scientist~\cite{google-co-scientist} and Claude Science~\cite{anthropic2026claudescience} are recent agentic systems that assist scientists with literature synthesis, data analysis and visualization, and manuscript drafting. In Claude Science, data analysis and visualization code runs on the user's infrastructure (including locally, over SSH, or via job schedulers), and provenance records are generated for the code, environment, and dependencies of all outputs. Unlike CURATE, these systems do not allow the generation of multi-step workflows targeting existing WMS, do not provide a way to reuse or share validated workflow components, and are locked-in to specific vendors.

\textbf{Connecting Agents to Cloud and HPC}: A growing body of work connects agents to cloud and HPC resources. Ma et al.~\cite{LangChain-Parsl} modified LangChain to allow agents to call functions scheduled on HPC resources using Parsl~\cite{Parsl}. In Yildiz and Peterka~\cite{do-llms-speak}, LLMs are evaluated on their ability to annotate, configure, and translate WMS files, demonstrating the potential for workflow generation and management. Looking beyond individual systems, Shin et al.~\cite{revolution} provide a broader overview of emerging workflow systems, arguing that workflows will evolve along two dimensions: composition (static to swarm) and intelligence (static to intelligent). Rather than connecting agents directly to distributed computing resources, \approach uses agents to generate workflow executables that are reproducible and can be deployed using legacy WMS.

\section{Architecture}
\label{sec:Architecture}

\subsection{Workflow Model}
Workflows are composed of nodes and edges, where each node represents a computational task to be completed, and edges define control dependencies. The underlying WMS determines the type of these nodes. We refer to the implementation of these nodes as \emph{modules}, which may be functions, scripts, or other types of executables with inputs and outputs. For our proposed architecture, task precedence must form a directed acyclic graph (DAG) and be known ahead of runtime scheduling. In Section~\ref{sec:prototype}, we present an implementation that, per prior workflow terminology~\cite{ornl-terminology}, targets a WMS with function-granularity, task-driven, file-transported workflows, where nodes/modules are implemented as Python functions.

\subsection{Overview}
Our proposed architecture (Fig~\ref {fig:curate-architecture}) consists of 4 agents: a Workflow Composition Agent (WCA), a Module Candidate Agent (MCA), a Module Generation Agent (MGA), and a Workflow Deployment Agent (WDA). In addition, the architecture includes a catalog for reusing workflow modules. Catalog reuse is enabled through \emph{source resolution}, which, in the context of \approach, refers to the process by which the MCA determines a node's module source based on its natural-language description in the skeleton. The resolved module source can be one of the following: 1) the catalog (if a match is found for reuse/adaptation); 2) a user-provided module; 3) a previous iteration of the workflow; or 4) a module to be generated by the MGA (if a match is not found). While the catalog itself can be extended as agentic memory~\cite{A-MEM}, the current prototype implements it as passive storage. Workflow generation begins with the user providing a high-level workflow description to the WCA -- a \emph{sketch} -- in natural language, and the system follows a sketch $\rightarrow$ reify $\rightarrow$ test $\rightarrow$ deploy set of stages, with the possibility of iterative HITL refining.

\subsection{Agent Descriptions}
\textbf{Workflow Composition Agent (WCA)}: The WCA controls the structure of the workflow. Based on the user's request, it generates a high-level skeleton of the workflow with natural-language descriptions of each node, providing an overview of the precedence of each step. This intermediate representation is presented to the user for verification via a user interface, such as an interactive graphical user interface (GUI) or terminal user interface (TUI), and the user may provide feedback. This check enables fine-grained control over task decomposition and allows users to catch errors early, before the code generation process begins. After the skeleton is created, it delegates source resolution to the MCA. Finally, the WCA resolves skeleton precedence edges with concrete inputs and outputs.

\textbf{Module Candidate Agent (MCA)}: The MCA resolves sources for skeleton tasks. When the WCA requests candidates from the MCA, it provides a high-level skeleton of the workflow. The MCA queries the catalog to see what existing modules are available. For each action in the skeleton, it maps it to a source. The following resolutions are available: freshly generated, from the catalog, a previous workflow revision, or a user-provided module. To make these classifications, the MCA is provided with additional workflow context, including the user's initial request, a description of the workflow management system, and the user's previous revisions. This agent enables users to reuse previously verified modules, either from their own previous runs or from other trusted users.

\textbf{Module Generation Agent (MGA)}: The MGA is tasked with implementing  workflow modules. It uses a coding agent to generate or adapt modules and to test them locally in a scratch directory, skipping nodes that are reused from the catalog or previous revisions. The MGA must ensure that each implemented module's I/O matches the WCA skeleton.

\textbf{Workflow Deployment Agent (WDA)}: The WDA deploys and continually monitors the workflow and its uploaded artifacts. After the workflow execution completes, the WDA spawns a post-run chatbot agent. This agent provides the user with natural language context about the run, including any failures that occur. The user can chat with this agent to ask questions about the run, and the agent invokes Model Context Protocol (MCP)~\cite{mcp-spec} tools to download artifacts and view logs before providing a natural-language response. This agent can also provide revision suggestions, including pointing out incorrect configurations, missing files, or genuine code issues, and ask the user for approval to revise.

\subsection{Catalog}
Cataloging serves as the basis for reusing workflow modules. When a user finishes composing a workflow, they are prompted to store it in the catalog. Depending on the target WMS, catalogs may consist of functions, ad hoc scripts, or, in future work with emerging workflow systems, agents with access to scientific instruments~\cite{revolution, academy}. During source resolution, the MCA queries the catalog and reuses entries where appropriate. Catalog entries may be stored locally on the user's file system or on cloud or remote storage. Users can share these catalogs by copying entries from other trusted users, serving as a foundation for future work that supports workflow reusability, discovery, and interoperability.

\subsection{User Supplied Module Reuse}
When the user chooses to provide an existing module, the MCA will determine whether to use it verbatim or adapt it. This selection is based on whether the module itself needs to change and excludes incompatible formatting for the underlying WMS. If the MCA determines to include a user-provided module verbatim, the MGA wraps it with I/O and the appropriate annotations, but does not change the module.

\subsection{Resource Isolation}
To mitigate the risk of agent-generated actions causing harm to user resources or data, or interfering with other processes, our proposed architecture ideally targets runtimes in which individual tasks can be sandboxed to isolated environments, such as containers or virtual machines. While these environments provides a foundation for the isolated execution of CURATE modules, care must also be taken to enforce isolation of state committed to centralized storage to prevent the propagation of side effects through file inputs/outputs that break downstream data dependencies. Future work will seek to incorporate new and existing abstractions for resource isolation into the AI-assisted workflow development process, from workflow definition to runtime environment.

\section{Prototype}
\label{sec:prototype}


We implemented an initial open-source~\cite{github-curate} prototype in  6700 lines of Python, using LangGraph for agent coordination and Claude Opus 4.8 as our underlying LLM. This prototype represents a minimal implementation of our proposed architecture, and as such, remains feasibility-focused. The MGA uses the Claude Agent SDK to generate function implementations; however, the prototype is extensible to support open-source coding agents, such as Codex, and open-source LLMs, with minimal changes. For deployment, we chose to target FaaSr~\cite{FaaSr}, a serverless FaaS WMS (originally targeting R, but since extended to Python), for the following reasons. First, FaaSr uses a schema-defined, task-driven, function-granularity DAG workflow model, with the only dynamicity being conditional branches, meaning all workflow structure and tasks are known ahead of time. The result is a WCA skeleton that can be represented as a simple DAG of functions. Second, serverless simplifies deployment, allowing our WDA to invoke the workflow without complex resource management. Finally, serverless deployment provides isolation via containerized execution of modules. The prototype exposes a command-line interface (CLI) app with a TUI viewer for code review and workflow monitoring. Since in serverless FaaS a module is expressed as a function, we use the terms \emph{module} and \emph{function} interchangeably in the following discussion.

\begin{figure}[t]
\centering
\includegraphics[width=\columnwidth]{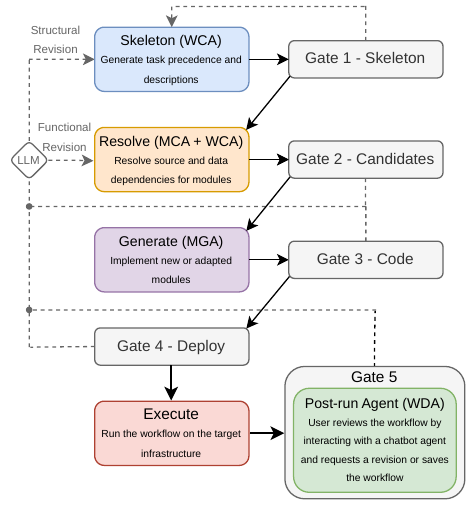}
\caption{A visualization of the 9-stage LangGraph pipeline, with 5 HITL gates for workflow revision. An LLM is used to determine whether revisions require structural changes or only affect functions already present in the skeleton.}
\label{fig:hitl-prototype}
\end{figure}

\subsection{Pipeline and HITL Gates}

The LangGraph pipeline comprises 9 stages and 5 HITL gates, where users can provide feedback for revision, as shown in Figure~\ref{fig:hitl-prototype}. At each gate, the user must approve the system's output and may provide optional revisions before moving on. 

\textbf{Skeleton and Gate 1}: The WCA generates a candidate skeleton that outlines the workflow's structure and high-level steps in natural language as sketched by the user. At this stage, the user can provide their own modules for reuse and context such as PDF documents, example I/O, or markdown files.

\textbf{Resolution and Gate 2}: After the user approves the skeleton at gate 1, the WCA fills in descriptions for the input and output edges between functions. Then, control shifts to the MCA, which determines an appropriate source for each function in the skeleton. First, it retrieves a list of potential functions from the catalog. Then it uses an LLM call to determine whether each node in the workflow should be resolved to the catalog, a previous iteration, or generated. If it chooses an existing function, it also determines whether it needs to be adapted to meet the function's requirements or can be reused verbatim. Then it returns its selections to the WCA, and the WCA fills in concrete dataflow (file names) between nodes and provides a list of likely needed PyPI dependencies. The user is then prompted to approve the filled-in (but not fully reified) workflow at gate 2.

\textbf{Generation and Gate 3}: Next, the MGA performs a topological sort of the skeleton and traverses it sequentially to generate each workflow function in its own scratch directory. The directory contains a \texttt{CONTEXT.md} file detailing shared context about the workflow, the user prompt, data integrity requirements, and the local FaaSr testing stubs, along with a per-function file specifying each function's description, signature, input/output files and dataflow, dependencies, and secret names (e.g., API key names), if any. Due to the topological generation ordering, the MGA sees any upstream-generated or cataloged functions. A single coding agent session is created using the Claude Agent SDK with access to this directory to generate one function per turn of the session, validating each function against the local stubs before moving on to the next. Once each function has been implemented, the user is shown a display at gate 3 where they can scroll through the code and the workflow file before approval.

\textbf{Execution and Gate 4}: At gate 4, the system requests approval to deploy the workflow, and it is deployed using FaaSr to GitHub Actions. Note that, although our prototype targets GitHub Actions, it is trivial to target other platforms using FaaSr (AWS Lambda, OpenWhisk, SLURM, Google Cloud Platform, and Kubernetes).

\textbf{Post-run agent and Gate 5}: Finally, if the deployment is successful, the post-run WDA agent serves as the final HITL gate. It provides the user with a natural-language summary of the workflow run and its artifacts. It exposes a chat interface that allows the user to ask interactive questions about the run, revise the workflow, commit their functions to the catalog for future reuse, and save the workflow for later deployment or adaptation. Upon execution failure, the WDA reviews the workflow artifacts and logs and advises the user on a recommended adaptation to submit for revision.

\subsection{Revision}
Upon revision, the system uses an LLM to classify whether the requested changes are structural or functional. Structural changes modify the underlying DAG, whereas functional changes modify functions already in the skeleton. Note that structural changes can only be made at skeleton creation, so structural revisions loop back to the skeleton step, while functional changes loop back to the resolution step.

\subsection{Function Generation}
For function generation, we use the Claude Agent SDK with Opus 4.8. When the MGA is deployed, the WCA provides the skeleton DAG with one of four labels for each function: \texttt{[New]}, \texttt{[Catalog]}, \texttt{[Cache]}, or \texttt{[User Provided Module]}; an optional \texttt{[Adapt]} label tells the MGA to use an existing function as context but adapt it to the user's requirements. Functions labeled \texttt{[Cache]} were implemented on an earlier revision cycle. Nodes labeled \texttt{[Catalog]} or \texttt{[Cache]} are skipped by the MGA unless they carry an \texttt{[Adapt]} label. For \texttt{[User Provided Module]} without \texttt{[Adapt]}, the MGA wraps the existing module with FaaSr I/O stubs for persistent data retrieval/storage from/to S3 cloud bucket(s) without changing the module's code, constructs synthetic data, and performs local parity checks of the wrapped function against the original. For \texttt{[New]} or \texttt{[Adapt]} functions, the MGA generates new code, in the latter case with the original function as context. The MGA is instructed not to return an implementation until it has tested the function using real data where available (produced by upstream functions), publicly available data (via an API), or local synthetic data when neither is available. Local testing is enabled by FaaSr stubs that mirror the actual FaaSr API. The FaaSr workflow JSON itself is created deterministically by means of a tool (not an agent) that translates the skeleton to the schema, and any required dependencies the MGA discovers beyond the WCA's list are written to a manifest and merged into the workflow configuration.

\subsection{Catalog Implementation and Search}

We implement our catalog on disk, where each entry comprises the
following fields: a unique \textbf{ID}; \textbf{name};
natural-language \textbf{description}; \textbf{inputs} and
\textbf{outputs}; PyPI \textbf{dependencies}; the names (not values) of \textbf{secrets}, if applicable;
\textbf{rank}; \textbf{keywords}; \textbf{provenance} (the workflow
from which the entry originates); and a \textbf{usage count}. To index the catalog, we use BM25~\cite{bm25} on the catalog entry keywords, using the tokenized natural-language description of each function provided by the WCA's skeleton as our search query. This catalog represents a minimal implementation; future work will seek to increase the scalability, fidelity, and correctness of component reuse and make catalogs compliant with FAIR principles.

\section{Experimental Results}
\label{sec:results}


To evaluate our prototype, we derived 4 experiments from SeBS-Flow~\cite{sebs-flow}, a serverless workflow benchmark suite, as well as 2 real-world examples automating and scaling a workflow that uses an existing Anaerobic Digestion Model (ADM1) implemented in Python (PyADM1). The ADM1 model is used in practice to size anaerobic digesters and forecast biogas production under varying conditions, though applying it to a specific facility requires extensive manual data reconciliation given the structure, frequency, quality, and accuracy of real-world data. The SeBS-Flow studies show the prototype system's ability to fully generate code and deploy a benchmark suite starting from a natural language prompt, while the PyADM1 studies demonstrate the system's ability to incorporate existing, unmodified model code with generated modules for data preprocessing, concurrent execution, summarization, and visualization.

We first recreate three SeBS-Flow workflows fully end-to-end: machine learning, MapReduce, and video analysis. Prompts used in the experiments are shown in the gray text blocks. For experiments 1-3, bolded text was taken directly from~\cite{sebs-flow}, and bulleted entries are the additional context added to ensure the full implementation can be recreated. These initial experiments demonstrate the system's ability to generate small workflows fully end-to-end while remaining faithful to strict user specifications. Experiment 4 demonstrates the system's ability to reuse cataloged modules and append additional MGA-implemented modules. Then, in experiment 5, we provide a description of a workflow that processes ``raw'' data to conform to PyADM1 inputs and execute the model, where the raw data carries faults representative of wastewater treatment plant: laboratory analytes measured weekly instead of the 15-minute frequency required by the simulation, instrumentation spikes as outliers, and units reported in the American convention (MGD, degrees Fahrenheit, mg/L) rather than metric as required by the model. In experiment 6, we extend experiment 5 to incorporate concurrent execution of a parameter sweep of the model across 20 initial states that vary the solids retention time (SRT). SRT is an operational parameter known to govern both digester sizing and methane production, but is not an explicit model input. For experiments 1-4, we compare our results against a ground truth. In 1-3, the ground truth is the original SeBS-Flow outputs; in 4, we manually calculate the correct word counts from the PDF. In experiments 5 and 6, we rely on domain expert knowledge to verify the behavior of data reconciliation and that model outputs respond to SRT in the expected directions. For brevity, we omit the original prompts from \cite{sebs-flow} for experiments 2 and 3, as well as the file directory names.

\subsection{Experiments}

\begin{promptbox}[Experiment 1 (Machine Learning)]
\textbf{MachineLearning: This workload represents a typical training pipeline: It starts with gen generating a dataset, with the number of samples N and the number of features M as input. Then, we train K different classifiers Ci in parallel. We generate N = 500 samples and M = 1024 features, and train K = 2 classifiers: a Support Vector Machine, and a Random Forest, creating two concurrent functions.}

\begin{promptitems}
\item For the dataset generation, use make\_classification with n\_samples=500, n\_features=1024, n\_redundant=0, n\_clusters\_per\_class=2, weights=[0.9,0.1], flip\_y=0.1, and random\_state=123
\item To preprocess use StandardScaler train\_test\_split with test\_size=0.4, random\_state=123
\item For Random Forest, set max depth to 5 and the number of estimators to 10; do not do any seeding
\item SVM must use SVC with kernel=linear and C at 0.025
\item For accuracy, use clf.score
\end{promptitems}
\end{promptbox}

\begin{revisionbox}
\begin{promptitems}
\item \textbf{Gate 1:} train\_classifier should be split into two concurrent functions
\end{promptitems}
\end{revisionbox}

In experiment 1, we generate a workflow that creates a synthetic dataset, preprocesses it, and trains SVM and Random Forest (RF) models concurrently. To ensure enough context to reproduce the original results, we include the dataset generation, preprocessing, and hyperparameter configurations in the prompt. In the initial skeleton, the WCA combined the two training functions into a single ranked function. After revision, train\_classifier was split into two functions (SVM and RF). Because both the original and the generated RF implementations were unseeded, we ran 60 invocations of both the original and generated workflows. The accuracy of SVM was identical, and the generated RF had an average accuracy of $0.850 \pm 0.002$, compared with the original's $0.851 \pm 0.006$.

\begin{promptbox}[Experiment 2 (MapReduce)]
\textbf{See section 5 of~\cite{sebs-flow} for the full prompt}

\begin{promptitems}
\item For the input text generation, use [cat, dog, bird, horse, pig], each distributed evenly and shuffled randomly
\item Aside from dataset generation, ensure MapReduce specific functions are fully generalizable to vocabularies of any size and frequency
\end{promptitems}
\end{promptbox}

In experiment 2, we generate a workflow that creates a list of words and runs a MapReduce algorithm split into four steps: Split $\rightarrow$ Map $\rightarrow$ Shuffle $\rightarrow$ Reduce. We provide information on the input data set so that the MGA can recreate the results of our SeBS-Flow baseline. The word counts from this workflow matched the SeBS-Flow ground truth verbatim.

\begin{promptbox}[Experiment 3 (Video Analysis)]
\textbf{See section 5 of~\cite{sebs-flow} for the full prompt}

\begin{promptitems}
\item Model: Faster R-CNN ResNet-50 COCO. Files frozen\_inference\_graph.pb and faster\_rcnn\_resnet50\_coco\_2018\_01\_28.pbtxt, loaded with OpenCV cv2.dnn
\item Three external inputs: frozen\_inference\_graph.pb, faster\_rcnn\_resnet50\_coco\_2018\_01\_28.pbtxt, and video\_small.mp4
\item Output: each detection's \{label, class, score\}, keeping score > 0.5; place in VideoAnalysis/
\item Provide the label for the object from the COCO 80 dataset using class\_id directly (zero-indexed, no offset; e.g. 0=person)
\end{promptitems}
\end{promptbox}

In experiment 3, we generate a workflow that takes as input a frozen inference graph, configurations for a faster R-CNN model trained on COCO, and a highway video, and classifies the objects in the video. In SeBS-Flow, the results were flattened to include only the outputs of a \emph{single} frame; we fixed this manually. For evaluation, we collapsed the correctness classification to whether the highest-probability object detected matched the ground truth for each frame (a car in all cases). The \approach workflow correctly identified the cars and did not produce any runtime errors. 

\begin{promptbox}[Experiment 4 (PDF + MapReduce + Visualize)]
Extract the words from words.pdf, split the word dataset, run MapReduce, visualize the output

\end{promptbox}

\begin{revisionbox}
\begin{promptitems}
\item \textbf{Gate 1:} split MapReduce into map, shuffle, reduce
\item \textbf{Gate 2:} reuse split verbatim and use the original rank for reduce and map
\end{promptitems}
\end{revisionbox}

In experiment 4, we reuse a subgraph of the previous MapReduce workflow, and extend it to: 1) use inputs from a PDF file rather than synthetic data, and 2) add a visualization function. The result is 4 reused functions (split, map, shuffle, reduce) and 2 new generated functions (extract\_words and visualize\_output). For our new input, we use a PDF with 5 words randomly distributed between 100 and 1500 occurrences. Note that we use the same vocabulary size as in the original (5), since the SeBS-Flow description ties the parallelism of the reducers to the vocabulary size, and structural revisions are beyond the scope of this experiment. The expected output matched the actual word counts, and the visualization (a histogram of the 5 words) was well formatted and clear.

\begin{promptbox}[Experiment 5 (PyADM1)]
Build a 7-step workflow that cleans raw digester influent data and runs a PyADM1 simulation on it.

1. convert-units: Reverse field units back to ADM1 units:
Q MGD$\rightarrow$m3/d, T (F)$\rightarrow$T (C) (convert AND rename the column),
divide ALL 22 COD columns by 1000 -- both solubles (S\_*) and
particulates (X\_ch, X\_pr, X\_li, X\_I, \ldots). Leave only S\_IC, S\_IN, S\_cation, S\_anion untouched. \\
2. fill-gaps: Forward-fill missing S\_cation and S\_anion values. \\
3. remove-spikes: Detect and replace isolated sensor spikes in BOTH Q and T with the local median. \\
4. interpolate: Treat the 26 weekly-held lab columns as weekly samples and interpolate to 15-min. \\
5. validate: Check the cleaned influent and initial-state files; write validation\_report.csv. \\
6. pyadm1: Run the PyADM1 simulation on the cleaned influent + initial state; write dynamic\_out.csv. \\
7. visualize: Plot the simulation outputs; write simulation\_plots.png.
\end{promptbox}
\begin{revisionbox}

\begin{promptitems}
\item \textbf{Gate 2}: the folder should be PyADM1-orig/
\end{promptitems}
\end{revisionbox}

In experiment 5, the workflow preprocesses ``raw'' influent data to conform to PyADM1 inputs, runs the model, and visualizes the results. We generate a synthetic ``raw'' influent input from the dataset distributed with PyADM1 by introducing changes that capture typical issues with raw data that require cleaning: unit differences, data gaps, and outliers. First, we converted the units of flow (Q, $\mathrm{m}^3/\mathrm{day}$ to MGD), temperature (T, C to F), and 22 concentration features ($\mathrm{kg/m}^3$ to $\mathrm{mg/L}$). Then, we applied weekly holds to the 26 laboratory features by freezing the first recorded value of each week. Next, we added gaps of 3--7 days in three places to S\_cation and S\_anion. Finally, we added implausible spikes to T and Q in four places each, simulating erroneous sensor readings. The CURATE-generated workflow was able to address most of the changes in the 5 preprocessing stages before running PyADM1, except the T spikes. T is hard-coded in PyADM1, so the MGA chose to skip the T spike fixes. Interestingly, the WDA caught this and suggested fixing the T spikes and modifying the PyADM1 model, but we chose to forgo this revision to keep the model unmodified. Note that, although the PyADM1 model is reused, the MGA still needed to wrap it with I/O and check parity with the original, driving up token usage.

\begin{promptbox}[Experiment 6 (PyADM1 with concurrency)]
\textbf{Same prompt as experiment 5, except for the following changes in steps 5 and 6:}\\
\\  
5. vary-inputs: create 20 inputs derived from the digester\_initial that vary SRT\\
6. pyadm1 (20x): Run the PyADM1 simulation on the cleaned influent + initial state; write dynamic\_out.csv.

\end{promptbox}

\begin{revisionbox}
\begin{promptitems}
\item \textbf{Gate 5 (WDA suggestion)}: In the \texttt{pyadm1} function, make the
simulation honor the per-rank SRT from
\texttt{digester\_initial.csv} - read the SRT value and set the
digester feed/flow so that hydraulic (and solids) retention time
equals SRT (e.g.\ set \texttt{q\_ad = V\_liq / SRT} instead of the
hard-coded \texttt{q\_ad = 178.4674}), so that the 20 ranked runs
produce distinct steady states across the 10-60 day sweep.
\end{promptitems}
\end{revisionbox}

\begin{table}[t]
\centering
\caption{Experimental Results}
\label{tab:correctness}
\small
\setlength{\tabcolsep}{0pt}
\renewcommand{\arraystretch}{1.15}
\newcommand{\none}{--}
\newcommand{\vgt}{\textcolor{green!55!black}{\cmark}\textsuperscript{\,\textcolor{green!55!black}{\textsc{gt}}}}
\newcommand{\vhv}{\textcolor{green!55!black}{\cmark}\textsuperscript{\,\textcolor{green!55!black}{\textsc{hv}}}}
\newcommand{\vev}{\textcolor{green!55!black}{\cmark}\textsuperscript{\,\textcolor{green!55!black}{\textsc{ev}}}}
\newcommand{\vzero}{\textcolor{red!80!black}{\xmark}\textsuperscript{\,\textcolor{black!60}{\textsc{0/3}}}}
\newcommand{\vone}{\textcolor{orange!90!black}{\cmark}\textsuperscript{\,\textcolor{black!60}{\textsc{1/3}}}}
\newcommand{\vtwo}{\textcolor{orange!90!black}{\cmark}\textsuperscript{\,\textcolor{black!60}{\textsc{2/3}}}}
\newcommand{\vthree}{\textcolor{green!55!black}{\cmark}\textsuperscript{\,\textcolor{black!60}{\textsc{3/3}}}}
\begin{tabular*}{\columnwidth}{@{\extracolsep{\fill}}lccccc c@{}}
\toprule
 & \multicolumn{5}{c}{\textbf{\approach}} & \textbf{Baseline} \\
\cmidrule(lr){2-6} \cmidrule(l){7-7}
\textbf{Exp.}
 & \shortstack{\textbf{Fn}\\\textbf{Gen.}}
 & \shortstack{\textbf{Fn}\\\textbf{Reused}}
 & \shortstack{\textbf{Fn}\\\textbf{Adapt.}}
 & \textbf{Rev.}
 & \textbf{Success}
 & \textbf{Success} \\
\midrule
E1 & 4 & 0 & 0 & G1     & \vgt & \vthree \\
E2 & 5 & 0 & 0 & \none  & \vgt & \vone   \\
E3 & 3 & 0 & 0 & \none  & \vgt & \vzero  \\
E4 & 2 & 4     & 0 & G1, G2 & \vgt & \vone   \\
E5 & 6 & 1 & 0     & G2     & \vev & \vone   \\
E6 & 1 & 4     & 2     & G5     & \vev & \vthree \\
\bottomrule
\end{tabular*}

\vspace{2pt}
\begin{minipage}[t]{\columnwidth}
\scriptsize
\textbf{Fn Gen}: number of functions generated by the MGA;
\textbf{Fn Reused}: number of functions reused from the catalog;
\textbf{Fn Adapt}: number of existing functions adapted to the prompt;
\textbf{Rev}: gates where revision occurred;
\textbf{Success}: success of the experiment (free from runtime errors and correct outputs) -- for \approach,
GT: ground truth, EV: expert verified; for the baseline,
the number of correct (and error-free) runs out of three compared to the verified \approach outputs
\end{minipage}
\end{table}

In experiment 6, we scaled the PyADM1 workflow to 20 concurrent invocations to perform a parameter sweep of SRT, which is controlled by flow into the digester and digester volume. We introduced a request in the prompt to vary inputs, modifying the preprocessed influent to use different SRT values. Upon deploying the workflow and reviewing outputs, the WDA identified an issue: q\_ad (the feed flow) was a hard-coded value in PyADM1, meaning that the SRT would not change despite the varied input. The WDA made a revision suggestion to PyADM1, which the HITL approved; after the revision, the workflow produced the correct results.

{\bf Baseline:} We compare our approach to a baseline that only uses the coding agent that implements the MGA: a plain Opus 4.8 Claude Code instance spawned using the Claude Agent SDK with the following context:

\begin{promptbox}[Baseline Context]
\begin{promptitems}
\item \url{http://faasr.io}
\item Web search
\item An example workflow JSON
\item A copy of the FaaSr workflow schema
\item An explanation of where to place external dependencies in the JSON
\item A list of available FaaSr API calls
\item The original experiment prompt
\item PyADM1 (for E5-6)
\item "Create a FaaSr workflow based on the user prompt. Your final output should be a FaaSr JSON file and functions written in Python"
\end{promptitems}
\end{promptbox}

To capture response variance with no room for HITL revision, we ran each baseline experiment 3 times. If any of the 3 runs succeeds (no runtime errors and correct outputs), the experiment is a success. Failures can stem from runtime errors or incorrect output. Every run uses the same context, except for E5-6, where PyADM1 and the raw inputs are provided. In many cases, there were runtime errors related to FaaSr that were not knowable from the provided context alone. For example, in some cases, the baseline agent mapped action names to dependencies instead of function names, or tried to split the action's code into multiple files. One of the FaaSr S3 API operations had an incorrectly documented argument name. In all such cases, we manually corrected the issue. We compare the results of our baseline to the ground truth or expert-verified \approach outputs.

\section{Discussion}
\label{sec:discussion}

\begin{figure}[t]
\includegraphics[width=\columnwidth]{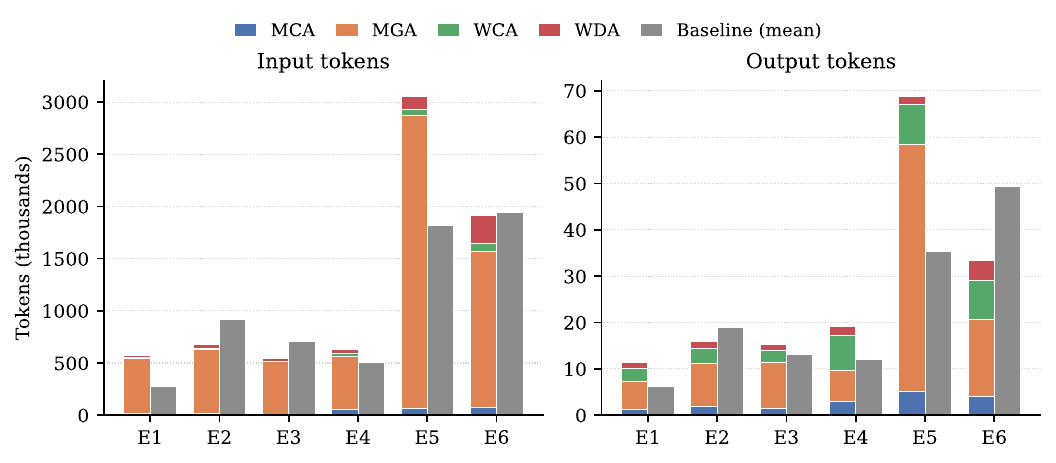}
\caption{Token usage per experiment (E1--E6). For each experiment, the left (colored, stacked) bar is the sum over \approach agents (MCA, MGA, WCA, WDA), and the right (grey) bar is the Claude Code baseline.}
\label{fig:tokens}
\end{figure}

Our \approach prototype demonstrates the \textit{feasibility} of a multi-agent workflow generation system, incorporating existing module reuse and HITL feedback revisions. Compared to a single-agent baseline, token usage (Figure~\ref{fig:tokens}) is higher in some cases; however, the baseline showed significant non-determinism in output correctness and, in many cases, did not produce a correct run and required manual revisions. In contrast with \approach, no such revisions were required aside from the natural language HITL feedback built into the system. \approach enables module reuse, which can improve the reliability of generated workflows by reusing verified modules. While the potential for decreasing token costs from reuse has been observed in some cases, further experiments are warranted - e.g. between E5 and E6, token usage nearly halved despite increased workflow complexity, while also requiring less HITL oversight of the preprocessing pipeline previously verified in E5. In E6, the post-run WDA successfully caught an issue with the PyADM1 model (where a parameter used to vary SRT was hard-coded), suggested a revision for HITL approval, and routed it back to the WCA/MCA for module resolution, highlighting a  benefit of the multi-agent integration. Table~\ref{tab:correctness} shows the results of each of the experiments, including the baseline results after corrections were applied to errors that were not reasonably knowable from the provided context. Finally, unlike the baseline, our prototype automates the deployment of workflows to a reproducible WMS.

While initial results are promising, our results have limitations. First, the Claude Code baseline did not include HITL revision at each step of the workflow-building process. We chose this comparison because Claude Code does not have a built-in notion of workflow creation, and it does not follow the same DAG sketch $\rightarrow$ reify $\rightarrow$ test model. Additionally, SeBS-Flow was originally designed as a performance benchmark rather than a scientific workflow suite; nonetheless, it provided a valuable reference for validating generated workflows. The ADM1 workflow tested is limited to initial data preparation. The full end-to-end workflow entails many additional steps, including preprocessing to achieve mass balance, extensive feature engineering, sensitivity analysis, and validation under unique conditions. However, operationalizing these steps is expected to be reasonably achievable with \approach, assuming practicing engineers can explicitly define the assumptions that govern certain actions. This could save a significant number of engineer-hours for mechanistic model calibration and significantly improve model accessibility, reproducibility, and utilization.

\section{Conclusion}
\label{sec:conclusion}
We presented \approach, a human-in-the-loop multi-agent system that manages workflows across their entire lifecycle, from composition to deployment and broader reuse. We demonstrated the system's feasibility with an initial open-source~\cite{github-curate} Python prototype and 6 evaluation experiments. Our prototype faithfully recreated 3 workflows in the SeBS-Flow serverless benchmark suite, reused a MapReduce pipeline from the catalog to create a new workflow, and automated and scaled a non-trivial anaerobic digestion simulation pipeline with preprocessing and visualization. Directions for future work include extending the catalog design to support semantic match queries and mechanisms for sharing and reuse across workflow repositories; interactive user interfaces for visual and conversational workflow generation; extensive user studies across different disciplines; and policies and evaluations focused on resource isolation, such as managing erroneous I/O operations generated by the MGA. 

\section*{Acknowledgments}
This work was supported by the U.S. National Science Foundation (NSF) under awards OAC-2450241, OAC-2311124, CNS-2533773, and CNS-2310422. Any opinions, findings, and conclusions expressed herein are those of the authors and do not necessarily reflect the views of the NSF. The authors also thank Dr. Nathan Tallent of Pacific Northwest National Laboratory (PNNL) for his feedback.

\bibliographystyle{IEEEtran}
\bibliography{references}


\end{document}